\documentclass[aps,pra,showpacs,twocolumn,superscriptaddress]{revtex4-1}
\usepackage{amsmath}
\usepackage{amsfonts}
\usepackage{amssymb}
\usepackage{graphicx}
\usepackage{natbib}
\usepackage[colorlinks=true,linkcolor=blue,urlcolor=black,citecolor=blue]{hyperref}

\usepackage{ulem}

\begin{document}
\title{Higher-order solutions to non-Markovian quantum dynamics via hierarchical functional derivative}
\author{Da-Wei Luo}
\affiliation{Beijing Computational Science Research Center, Beijing 100094, China}
\author{Chi-Hang Lam}
\affiliation{Department of Applied Physics, Hong Kong Polytechnic University, Hung Hom, Hong Kong, China}
\author{Lian-Ao Wu}
\affiliation{Department of Theoretical Physics and History of Science, The Basque Country University (EHU/UPV), PO Box 644, 48080 Bilbao, Spain}
\affiliation{Ikerbasque, Basque Foundation for Science, 48011 Bilbao, Spain}
\author{Ting Yu}
\affiliation{Beijing Computational Science Research Center, Beijing 100094, China}
\affiliation{Center for Controlled Quantum Systems and Department of Physics and Engineering Physics, Stevens Institute of Technology, Hoboken, New Jersey 07030, USA}
\author{Hai-Qing Lin}
\affiliation{Beijing Computational Science Research Center, Beijing 100094, China}
\author{J. Q. You}
\affiliation{Beijing Computational Science Research Center, Beijing 100094, China}
\date{\today}

\begin{abstract}
Solving realistic quantum systems coupled to an environment is a challenging task. Here we develop a hierarchical functional derivative (HFD) approach for efficiently solving the non-Markovian quantum trajectories of an open quantum system embedded in a bosonic bath. An explicit expression for arbitrary order HFD equation is derived systematically. Moreover, it is found that for an analytically solvable model, this hierarchical equation naturally terminates at a given order and thus becomes exactly solvable. This HFD approach provides a systematic method to study the non-Markovian quantum dynamics of an open system coupled to a bosonic environment.
\end{abstract}

\maketitle

\section{Introduction}
The theory of open quantum systems~\cite{Breuer2002} has received great interest because environment-induced effects are important in a wide range of research topics such as quantum information~\cite{Nielsen2000} and quantum optics~\cite{Scully1997}. There were considerable studies involving environments modeled by either bosonic or fermionic baths (see, e.g., Refs.~\cite{Breuer2002,Diosi1998,Strunz1999,Yu1999,Gambetta2002,Jing2010_2012,Li2014,Suess2014,diosi2014,fermionic_Chen2013,Fermionic_Zhao2012}), as well as structured environments such as spin-chain baths~\cite{Quan2006_and_me}. Conventionally, Markov approximation has been extensively used because of its simplicity. Indeed, this is valid only when memory effects of the environment are negligible. However, this approximation becomes invalid when the system-environment coupling is strong or when the environment is structured~\cite{Breuer2002,Ma2014}. Therefore, non-Markovian environments have to be considered in explaining new experimental advances in quantum optics. Also, they must be considered in quantum information manipulations in which the environmental memory is utilized to control the entanglement dynamics~\cite{Bellomo2007}. Therefore, it is vital to have a non-Markovian description of the system's dynamics under the influence of the memory effects and the backaction of the environment. Actually, this has long been a challenging task and many theoretical approaches have been developed (see, e.g., Refs.~\cite{Breuer2002,Diosi1998,Strunz1999,Yu1999,Gambetta2002,Jing2010_2012,Li2014,Suess2014,
hpz,Li2011,Bassi2003,Breuer2004,Diosi1998,Piilo2009}). Among these approaches, the non-Markovian quantum state diffusion (QSD)~\cite{Diosi1998,Strunz1999,Yu1999} has been proven to be a powerful tool to study the quantum dynamics of the system and exact analytical results were derived for many interesting systems such as dissipative multi-level atoms~\cite{Jing2010_2012} and quantum Brownian motion~\cite{Diosi1998,Strunz2004} which was also analytically solved via a path-integral approach~\cite{hpz}. Quantum continuous measurement
employing the QSD technique was also studied~\cite{qsd_qm,qsd_qc}.

For most realistic open quantum system problems, it is almost impossible to find any useful analytical solutions.  Therefore, one has to develop numerical methods to study the quantum dynamics of the open systems. However, the application of the non-Markovian QSD is greatly hindered unless one can cast it to a numerically implementable time-local form. Recently, two hierarchical approaches have been proposed;
one is the stochastic differential equation (SDE) method~\cite{Li2014} based on a functional expansion of a system operator, and the other is an approach using a hierarchy of pure states (HOPS)~\cite{Suess2014} to calculate a related functional derivative. Apart from these two approaches, an earlier attempt at a perturbative solution of the non-Markovian QSD equation was based on hierarchical functional derivative (HFD)~\cite{Gambetta2002}.
However, because of its apparent complexity, the hierarchical equations presented there were implemented only up to the second order where the higher-order terms were approximated by a simplified operator.
Here we develop a systematic and efficient higher-order HFD approach to solving the non-Markovian quantum dynamics of an open system coupled to a bosonic environment.
Remarkably, a compact explicit expression for an arbitrary order hierarchical equation is derived. Moreover, it is found that for an analytically solvable model, the hierarchical equation naturally terminates at a given order, so it becomes exactly solvable. Thus, our method provides not only an approach for efficiently solving the non-Markovian quantum dynamics of an open system, but also  a systematic method for the exact solution of analytically tractable open systems.

The paper is organized as follows. In Sec.~\ref{sec_h}, we present our HFD method for studying the non-Markovian QSD. Then, we show the relationship of the HFD method to the recently developed SDE and HOPS method in Sec.~\ref{sec_rel_sde} and Sec.~\ref{sec_rel_hop}, respectively. Finally, discussion and conclusion are given in Sec.~\ref{sec_d_and_c}. Moreover, in Appendix~\ref{ax_s_gen}, we present a detailed derivation of the HFD equation for a general bath correlation function. The mathematical relationship between the SDE approach and our HFD method is explicitly demostrated in Appendix~\ref{ax_s_rel}.

\section{The HFD method for non-Markovian QSD}\label{sec_h}

We study a generic open system with the Hamiltonian~\cite{Diosi1998,Strunz1999}
\begin{equation}
	H=H_{s}+\sum_k \left(g_k L b_k ^\dagger+g_k^* L^\dagger b_k \right)+\sum_k \omega_k b_k ^\dagger b_k,
\end{equation}
where $H_{s}$ is the Hamiltonian of the system of interest, $L$ is the coupling operator called Lindblad operator, $b_k$ is the $k$th mode annihilation operator of the bosonic bath with a frequency $\omega_k$ and $g_k$ denotes the coupling strength between the system and the bosonic bath. The bath state can be specified by a set of complex numbers $\{z_k\}$ labeling the coherent state of all bath modes and the effect of the bath is characterized by the zero-temperature bath correlation function $\alpha(t,s)=\sum_k |g_k|^2 e^{-i \omega_k (t-s)}$. Defining $z_t^*\equiv -i\sum_k g_k^* z_k^* e^{i \omega_k t}$, one can interpret $z_k$ as a Gaussian random variable and $z_t^*$ becomes a Gaussian process with its statistical mean given by the bath correlation function $\alpha(t,s)= \langle\langle z_t z_s^* \rangle\rangle$.
The wave function $|\psi_{z^*}(t) \rangle=\langle z^*|\Psi_{\rm{tot}}(t)\rangle$, obtained by projecting the quantum state $|\Psi_{\rm{tot}}(t)\rangle$ of the total system onto the bath state $|z \rangle$, is called a quantum trajectory and obeys a linear QSD equation
\begin{equation}
	\frac{\partial}{\partial t}|\psi_{z^*}(t) \rangle= \left[-iH_{s}+Lz^*-L ^\dagger \bar{O}(t,z)\right]|\psi_{z^*}(t) \rangle,
\end{equation}
where  $O$ is an operator defined by the functional derivative $\frac{\delta}{\delta z_s^*}|\psi_{z^*}(t) \rangle=O(t,s,z^*)|\psi_{z^*}(t) \rangle$ and $\bar{O}(t,z^*)=\int_0^t \alpha(t,s) O(t,s,z^*)ds$ ~\cite{Diosi1998}. Non-Markovian master equations can also be obtained using this approach~\cite{Strunz2004,Yu1999}. For this linear QSD, $|\psi_{z^*}(t) \rangle$ is an unnormalized wave function, so it can be of different orders of magnitude at different evolution times. This gives rise to an inefficient Monte Carlo sampling in numerical calculations. Thus, one utilizes a non-linear QSD~\cite{Diosi1998} for the normalized state $|\tilde{\psi}_{\tilde{z}^*} \rangle=|\psi_{z^*}(t) \rangle/|||\psi_{z^*}(t) \rangle||$:
\begin{align}
	\frac{d|\tilde{\psi}_{\tilde{z}^*} \rangle}{dt}=&\left[-iH_s+\Delta_t(L)\tilde{z}^*_t-\Delta_t(L^\dagger)\bar{O}(t,\tilde{z}^*)\right. \nonumber\\
	&\left. + \langle\Delta_t(L^\dagger) \bar{O}(t,\tilde{z}^*)\rangle_t \right]|\tilde{\psi}_{\tilde{z}^*}(t)\rangle,\label{qsd_nl}
\end{align}
which is derived via the Girsanov transformation
$\tilde{z}^*_t=z_t^*+\int_0^t \alpha^*(t,s) \langle L ^\dagger \rangle_s ds$. Here $\Delta_t(L)\equiv L-\langle L \rangle_t$, and the reduced density operator $\rho_s(t)\equiv\mathrm{Tr}_{\rm{env}}|\Psi_{\rm{tot}} \rangle\langle \Psi_{\rm{tot}}|$ can be obtained from the ensemble average of the normalized quantum trajectories as $\rho_s(t)= \langle\langle|\tilde{\psi}_{\tilde{z}^*}(t) \rangle \langle\tilde{\psi}_{\tilde{z}}(t)|\rangle\rangle$.

The non-Markovian QSD equation~\eqref{qsd_nl} is exact, but the key challenge in solving it relies on the determination of the $O$ operator. It is difficult to analytically obtain the explicit expression of the $O$ operator except for a few simple models such as dissipative multi-level atoms~\cite{Jing2010_2012}, the quantum Brownian motion~\cite{Diosi1998} and dissipative multiple qubits~\cite{Jing2013}. Owing to the importance of open systems, efforts~\cite{Gambetta2002,Li2014,Suess2014} have been devoted to develop numerical methods for solving Eq.~\eqref{qsd_nl}. In Ref.~\cite{Gambetta2002}, an approach was proposed to calculate the functional derivative by defining a set of $\mathcal{Q}_k$ operators as
\begin{equation}
	\mathcal{Q}_k(t,\tilde{z}^*)=\int_0^t \alpha(t,s) \frac{\delta}{\delta \tilde{z}^*_s}\mathcal{Q}_{k-1}(t,\tilde{z}^*)ds,\label{fdef}
\end{equation}
where $\mathcal{Q}_0(t,\tilde{z}^*)=\bar{O}(t,\tilde{z}^*)$. Because of the complexity involved, an explicit hierarchical equation of motion for this set of operators was obtained in Ref.~\cite{Gambetta2002} only up to the second-order $\mathcal{Q}_2$ operator while the higher-order terms were approximated by a simplified operator. Below we give an explicit hierarchical equation of motion up to any high orders and also applicable to an arbitrary bath correlation function $\alpha(t,s)$. For this purpose, we include the time derivatives of $\alpha(t,s)$ and define
\begin{align}
	\mathcal{Q}_k^{(\mathbf{j})}(t,z^*)=&\mathcal{P}^{(j_k)}\mathcal{P}^{(j_{k-1})} \ldots \mathcal{P}^{(j_1)}\nonumber\\
	&\times \int_0^t ds_0 \alpha^{(j_0)}(t,s_0) O(t,s_0,z^*),
\end{align}
where $\mathbf{j}=\sum_i j_i \mathbf{e_i}\equiv (j_0,j_1,\ldots,j_k)$ with $\mathbf{e_i}$ being the $i$th unit vector, $\mathcal{P}^{(j)}=\int_0^t ds \alpha^{(j)}(t,s) \frac{\delta}{\delta z^*_s}$, and $\alpha^{(j)}(t,s)=\frac{\partial^j}{\partial t^j}\alpha(t,s)$. It is derived (see Appendix~\ref{ax_s_gen}) that this set of operators satisfy
\begin{align}
	\frac{\partial}{\partial t} &\mathcal{Q}_k^{(\mathbf{j})}(t,z^*)=\sum_{i=1}^k \alpha^{(j_i)}(0)\left[L,\mathcal{Q}_{k-1}^{(\mathcal{D}(\mathbf{j},i))}(t,z^*)\right] \nonumber\\
	&+\sum_{i=0}^k \mathcal{Q}_k^{(\mathbf{j} +\mathbf{e_i})}(t,z^*)-L ^\dagger \mathcal{Q}_{k+1}^{(0,\mathbf{j})}(t,z^*) \nonumber\\
	&+\left[-iH_{\rm sys} +Lz_t^* ,\mathcal{Q}_k^{(\mathbf{j})}(t,z^*)\right] +\delta_{0,k}\alpha^{(j)}(0)L\nonumber\\
	&-\sum_{i=0}^k\sum_\mathbf{c_i}\left[L ^\dagger \mathcal{Q}_i^{(0,\mathbf{c_i})}(t,z^*),\mathcal{Q}_{k-i}^{(j_0,\mathbf{\bar{c}_i})}(t,z^*)\right],\label{gen_a_qj}
\end{align}
where $\mathcal{Q}_k^{(\mathbf{0})}(t,\tilde{z}^*)=\mathcal{Q}_k(t,\tilde{z}^*)$, $\alpha^{(j)}(0)=\alpha^{(j)}(t,t)$, $\mathcal{D}(\mathbf{j},i)\equiv(j_0,\ldots j_{i-1},j_{i+1},\ldots,j_k)$ excludes $j_i$ in the vector $\mathbf{j}$, and $\sum_{\mathbf{c_i}}$ is the sum of $k!/[i!(k-i)!]$ terms which include all possible cases of choosing $i$ elements from a $k$-dimensional vector $(j_1,\ldots,j_k)$. Using functional expansions~\cite{Yu1999}, we can also immediately prove that $\mathcal{Q}_k^{(\mathbf{j})}(t,z^*)$ operators have such a symmetry that $\mathcal{Q}_k^{(j_0,j_1,\ldots,j_k)}(t,z^*)$ coincide for all permutations of $(j_1,j_2,\ldots,j_k)$. This enables us to greatly reduce the number of equations that should be solved, because we only need to calculate those $\mathcal{Q}_k^{(\mathbf{j})}(t,z^*)$ operators with normal-ordered $j_i$ ($j_1\leq j_2\leq \ldots \leq j_k$).

It is also worth pointing out that since the QSD equation for a finite-temperature bath with self-adjoint Lindblad operator $L$ (i.e., $L=L^{\dag}$) is formally the same as the zero-temperature one with a different finite-temperature bath correlation function $\alpha(t,s)$~\cite{Diosi1998}, so Eq.~\eqref{gen_a_qj} also applies to that finite-temperature case. Note that the finite-temperature case with a self-adjoint $L$ covers a large number of open-system problems, including the important spin-boson model without the rotating-wave approximation (see, e.g., \cite{Li2014}).

When modeling the structure of the bosonic bath, an important and widely used choice is the Lorentzian spectrum, which corresponds to an environmental noise $z_t$ of the Ornstein-Uhlenbeck type with autocorrelation $\alpha(t,s)=\Gamma \gamma \exp(- \gamma|t-s|)/2$. This kind of bosonic bath has been used to describe many interesting problems~\cite{Scully1997,Breuer2002}, and it is easy to observe a non-Markovian to Markovian crossover by just increasing $\gamma$. In addition, this can greatly simplify the hierarchical equations since $\alpha^{(j)}(t,s)=(-\gamma)^j \alpha(t,s)$ and $\mathcal{Q}_k^{(\mathbf{j})}=(-\gamma)^{J_s}\mathcal{Q}_k$, where $J_s=\sum_i j_i$. In this case, by utilizing our general result~\eqref{gen_a_qj}, a compact hierarchical equation for the  $\mathcal{Q}_k$ operators given by Eq.~\eqref{fdef} can be explicitly written as
\begin{align}
	\frac{\partial}{\partial t}& \mathcal{Q}_{k}(t,\tilde{z}^*)=k \alpha(0)\left[L,\mathcal{Q}_{k-1}(t,\tilde{z}^*)\right] +\delta_{0,k}\alpha(0)L\nonumber\\
	&-(k+1)\gamma \mathcal{Q}_{k}(t,\tilde{z}^*)+\left[-iH_s+ L\tilde{z}^*_t,\mathcal{Q}_{k}(t,\tilde{z}^*)\right]\label{fk_heq}\\
	& -L^\dagger \mathcal{Q}_{k+1}(t,\tilde{z}^*) -\sum_{i=0}^{k}C_i^k \left[L ^\dagger \mathcal{Q}_i(t,\tilde{z}^*),\mathcal{Q}_{k-i}(t,\tilde{z}^*)\right],\nonumber
\end{align}
where $C_i^k\equiv k!/[i!(k-i)!]$ is the binomial coefficient and the initial conditions are $\mathcal{Q}_k(0,\tilde{z}^*)=0$ for $k\geq 0$. This set of hierarchical equations can be numerically solved perturbatively if terminated at order $\mathcal{N}+1$ by putting either $\mathcal{Q}_{\mathcal{N}+1}(t,\tilde{z}^*)=0$ or $\mathcal{Q}_{\mathcal{N}+1}(t,\tilde{z}^*)=\int_0^t ds \alpha(t,s) \left[L,\mathcal{Q}_{\mathcal{N}}(t,\tilde{z}^*)\right]$ ~\cite{Gambetta2002}.

\begin{figure}
	\centering
	\includegraphics[scale=.4]{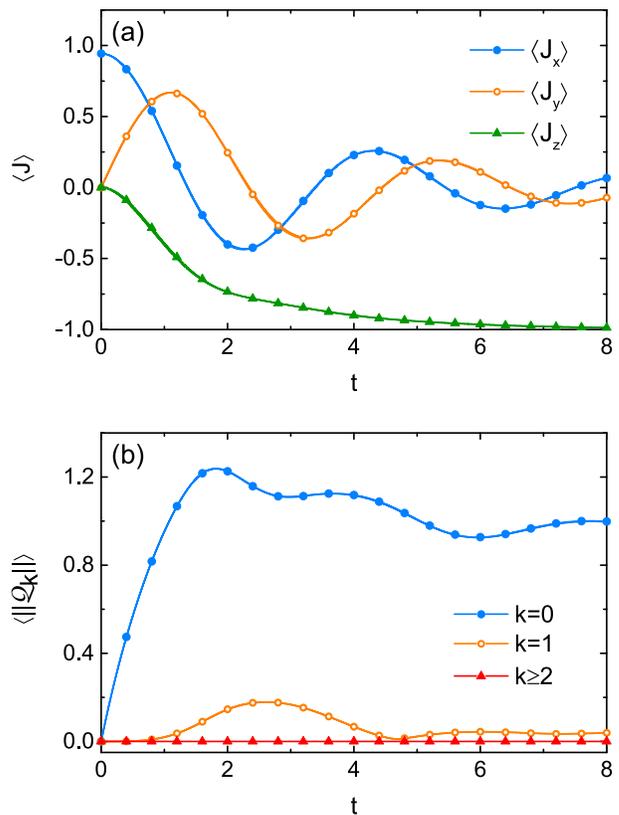}\\
	\caption{(Color online) (a)~Ensemble average values of the angular momentum $\langle J_{x}\rangle$, $\langle J_{y}\rangle$, and $\langle J_{z}\rangle$ versus time $t$. (b)~Ensemble average values of the trace norm of $\mathcal{Q}_0(t,\tilde{z}^*)$, $\mathcal{Q}_1(t,\tilde{z}^*)$, and $\mathcal{Q}_{k}(t,\tilde{z}^*)$, with $k\geq 2$, as a function of $t$. In both (a) and (b), we use $1000$ noise realizations and $\gamma=\Gamma=\omega=1$. The solid curves correspond to the analytic solutions; the solid, open circles and triangles correspond to the results obtained using our HFD approach.}\label{figu}
\end{figure}

Now we also explore the use of this approach as a systematic method for models with exact solutions. For open systems that are known to be exactly solvable by QSD, the number of expansion terms in
\begin{align}
	\bar{O}&(t,\tilde{z}^*)=\bar{O}^{(0)}(t)+\int_0^t\bar{O}^{(1)}(t,v_1)\tilde{z}^*_{v_{1}}dv_1 \nonumber\\
	 &+\int_0^t\int_0^t\bar{O}^{(2)}(t,v_1,v_2)\tilde{z}^*_{v_{1}}\tilde{z}^*_{v_{2}}dv_1dv_2+\ldots.\label{eq_funcexp_def}
\end{align}
must be finite~\cite{Yu1999}. Thus, $\bar{O}^{(n)}\equiv 0$ for all $n$ larger than a given finite integer  $N_c$. In this case, we can readily show that
\begin{align}
	\mathcal{Q}_{N_c}(t)=&N_c!\int_0^t \ldots \int_0^t \alpha(t,v_1) \ldots \alpha(t,v_{N_c}) \nonumber\\
	&\times \bar{O}^{(N_c)}(t,v_1,\ldots v_{N_c}) dv_1\ldots dv_{N_c},\label{fnc}
\end{align}
which is independent of the noise $\tilde{z}_t^*$. Therefore, its functional derivative with respect to $\tilde{z}_s^*$ is zero. From Eq.~\eqref{fdef}, it follows that $\mathcal{Q}_{k}=0$ for all $k\geq N_c+1$, so that the hierarchical equation naturally terminates and the approach becomes exact. This provides a useful and systematic method to deal with an open system with unknown properties by just implementing the hierarchical equation; if the hierarchical equation has a natural termination at a given order, the considered model is analytically solvable  and our results will be essentially accurate.

As an illustrative example, we consider an analytically solvable three-level system~\cite{Jing2010_2012}, with $H_{\rm sys}=\omega J_z$, and $L=J_-$. The functional expansion of its $O$ operator is only up to the first-order term, i.e., $N_c=1$. Thus, only $\mathcal{Q}_0(t,\tilde{z}^*)$ and $\mathcal{Q}_1(t)$ are involved in the hierarchical equation, and $\mathcal{Q}_k\equiv 0$ for $k\geq 2$. We numerically solve the hierarchical equation~\eqref{fk_heq} up to order $\mathcal{N}=10$. From Fig.~\ref{figu}(a), it can be seen that the numerically calculated ensemble averages $\langle J_i \rangle$, $i=x$, $y$ and $z$, agree well with the exactly solved results. Also, the trace norms  $||\mathcal{Q}_k||\equiv\mathrm{Tr}\left(\sqrt{\mathcal{Q}_k ^\dagger \mathcal{Q}_k}\right)$ are calculated in Fig.~\ref{figu}(b), which shows that for $k\geq 2$, the $\mathcal{Q}_k$ operators remain zero and the hierarchical equation naturally terminates at order $k=2$ (i.e., $N_c=1$), in full consistency with the analytical derivations.

\section{The relationship to the SDE method}\label{sec_rel_sde}

Recently, a numerical approach~\cite{Li2014} was developed to solve the non-Markovian QSD equation via a set of stochastic differential equations (SDE). A key part of the SDE formulation is the introduction of a $Q$ operator
\begin{align}
	Q_m^{(n)}(t,\tilde{z}^*)=&\int_0^t \ldots \int_0^t \alpha(t-v_1)\ldots\alpha(t-v_m)\tilde{z}^*_{v_{m+1}} \nonumber\\
	&\ldots\tilde{z}^*_{v_{n}}\bar{O}^{(n)}(t,v_1,\ldots,v_n)dv_1 \ldots dv_n,\label{qdef}
\end{align}
where $\bar{O}^{(n)}(t,v_1,\ldots,v_n)$ corresponds to the $n$th-order functional expansion term in Eq.~\eqref{eq_funcexp_def} and $\bar{O}(t,\tilde{z}^*)=\sum_{n=0}Q_0^{(n)}(t,\tilde{z}^*)$. For $m\neq 0$, $Q_m^{(n)}$ do not directly contribute to $\bar{O}(t,\tilde{z}^*)$ but form a set of hierarchical equations with $Q_0^{(n)}$ and need to be solved simultaneously.
Within this framework, one can calculate the quantum trajectory up to an arbitrary order of the environmental noise.

\begin{figure}
	\centering
	\includegraphics[scale=.4]{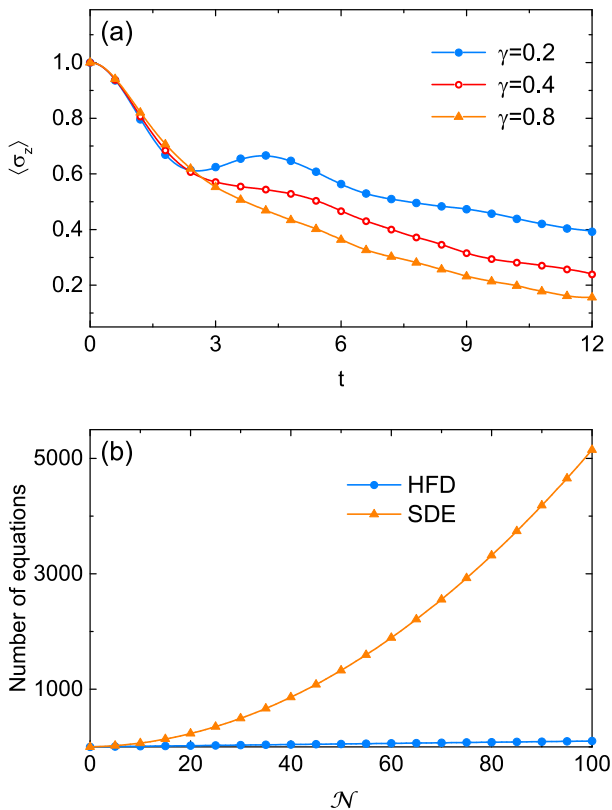}
	\caption{(Color online) (a)~Ensemble average $\langle \sigma_{z}\rangle$ versus time $t$ for $\gamma=0.2$, $0.4$ and $0.8$, obtained using the SDE approach (solid curves) and the HFD method (solid, open circles and triangles). Here $\mathcal{N}=30$, $\Gamma\gamma=0.2$, and $1000$ noise realizations are used. (b)~Total number of equations versus termination order $\mathcal{N}$, where the curve marked by triangles (solid circles) corresponds to the number of coupled equations of motion that should be simultaneously solved in the SDE (HFD) approach.}\label{figu_sbx}
\end{figure}

The spin-boson model without the rotating-wave approximation (RWA) was studied using this approach by considering an Ornstein-Uhlenbeck noise (see \cite{Li2014}). This is a typical example where the explicit form of the $O$ operator is unknown and the hierarchical approach can serve as a powerful numerical tool. In Fig.~\ref{figu_sbx} (a), we display the expectation value of the angular momentum $\langle \sigma_z\rangle$ as a function of time using both the SDE and our HFD approaches. Here we also use an Ornstein-Uhlenbeck noise, so as to directly compare with the SDE results. An excellent agreement is reached for the numerical results obtained by them. As a matter of fact, these two hierarchical methods are very closely related mathematically, and we prove in Appendix~\ref{ax_s_rel} that the operators $\mathcal{Q}_k$ and $Q_m^{(n)}$ are related by
\begin{align}
	\mathcal{Q}_k(t,\tilde{z}^*)=\sum_{n=k}^\mathcal{N} \frac{n!}{(n-k)!}Q_k^{(n)}(t,\tilde{z}^*).\label{fkq}
\end{align}
The key advantage of the HFD method over the SDE approach is that the HFD equation~\eqref{fk_heq} effectively groups the $Q_m^{(n)}$ operator according to Eq.~\eqref{fkq} and sums up their contributions. This makes the HFD method much more efficient than the SDE approach. In fact, for a given termination order $\mathcal{N}$, the SDE approach needs to simultaneously solve coupled equations for $Q_m^{(n)}$ where $n+m\leq \mathcal{N}$, resulting in a total number of $\frac{1}{2}(\mathcal{N}+2)(\mathcal{N}+1)$ equations. On the other hand, the HFD approach greatly reduces the total number of equations to $\mathcal{N}+1$. Figure~\ref{figu_sbx}(b) shows the total number of coupled differential equations that should be solved in both the HFD and the SDE approaches. The very high efficiency of the HFD method over the SDE approach is clearly seen when increasing $\mathcal{N}$.

\section{The relationship to the HOPS method}\label{sec_rel_hop}

Previously, almost all QSD approaches are focused on how to calculate the functional derivative associated with the $O$ operator. Recently, a hierarchy of pure states (HOPS) approach~\cite{Suess2014} was developed as a numerical tool which, instead of using the $O$ operator, introduces a set of pure states
\begin{align}
	|\psi_k(t) \rangle=\int_0^t \alpha(t,s)\frac{\delta}{\delta \tilde{z}^*_s}ds|\psi_{k-1}(t)\rangle,
\end{align}
where $|\psi_0(t) \rangle\equiv|\psi(t)\rangle$. In this approach, a set of hierarchical equations of motion was found for $|\psi_k (t)\rangle$. Its advantage is that the hierarchical equations deal with state vectors of size $\rm{dim}(\mathcal{H}_s)\times \mathbf{1}$ rather than the operators of size $\rm{dim}(\mathcal{H}_s)\times \rm{dim}(\mathcal{H}_s)$, where $\rm{dim}(\mathcal{H}_s)$ is the dimension of the system's Hilbert space. From the definition, it is easy to see that
\begin{align}
	|\psi_1(t) \rangle=&\mathcal{Q}_0(t,\tilde{z}^*)|\psi_0(t) \rangle, \nonumber\\
	|\psi_2(t) \rangle=&\mathcal{Q}_1(t,\tilde{z}^*)|\psi_0(t) \rangle+\mathcal{Q}_0(t,\tilde{z}^*)|\psi_1(t) \rangle, \nonumber\\
	|\psi_3(t) \rangle=&\mathcal{Q}_2(t,\tilde{z}^*)|\psi_0(t) \rangle+2\mathcal{Q}_1(t,\tilde{z}^*)|\psi_1(t) \rangle \nonumber\\
	&+\mathcal{Q}_0(t,\tilde{z}^*)|\psi_2(t) \rangle, ~~\ldots\ldots,\nonumber	
\end{align}
and in general,
\begin{equation}
	|\psi_k \rangle=\sum_{i=0}^{k-1}C_i^{k-1}\mathcal{Q}_i(t,\tilde{z}^*)|\psi_{k-i-1} \rangle. \label{hops_and_us}
\end{equation}
Therefore, it is also possible to formulate the HOPS approach using the $\mathcal{Q}_k$ operators in Eq.~\eqref{fdef}. As shown above, for an exactly solvable model, $\mathcal{Q}_i(t,\tilde{z}^*)=0$ for all $i\geq N_c+1$, so that the hierarchical equations for $\mathcal{Q}_i(t,\tilde{z}^*)$ terminate at the $N_c$th order. It is interesting to note that there are infinite number of hierarchical equations for $|\psi_k \rangle$ in the HOPS approach, but only $N_c+1$ nonzero operators $\mathcal{Q}_0(t,\tilde{z}^*),\mathcal{Q}_1(t,\tilde{z}^*),\dots,\mathcal{Q}_{N_c}(t,\tilde{z}^*)$ are needed to express all of these $|\psi_k \rangle$. This reflects the exact solvability of the considered model.

\section{Discussion and Conclusion}\label{sec_d_and_c}

Solving non-Markovian dynamics of an open quantum system has long been a challenge. The conventional master equation for the system's reduced density matrix is driven by a super-operator $\mathcal{K}$, which is given by the perturbation expansion with respect to the strength of system-bath interaction~\cite{Breuer2002}. Whereas the expansion could be done manually up to the orders beyond 2, it is difficult to find an automatic numerical way to obtain the higher-order $\mathcal{K}$ operator. As such, the conventional master equation is used in the weakly-coupled scenario. Likewise, the quantum state diffusion equation driven by an $O$ operator (which plays a role similar to $\mathcal{K}$) encounters the same problem. As a breakthrough, this work develops a systematic and efficient higher-order HFD approach for solving the non-Markovian quantum trajectories of an open system coupled to a bosonic environment. A compact explicit expression for an arbitrary order hierarchical equation of motion is derived and it can be efficiently implemented numerically. As a distinctive advantage of this method, while this hierarchical equation naturally terminates at a given order and becomes exactly solvable for an analytically solvable model, it provides a systematic perturbation for a generic open system irrespective of the existence of the time-local $O$ operator. Our HFD method applies to an arbitrary bath correlation function. Also, it can be naturally extended to
the case of a finite-temperature bath when the Lindblad operator in the interaction Hamiltonian is self-adjoint, including the important spin-boson model without the rotating-wave approximation.

\vspace{4mm}
\begin{acknowledgments}
This work is supported by the National Natural Science Foundation of China No.~91421102 and the National Basic Research Program of China No.~2014CB921401. C.H.L. is supported by HK GRF (Project No.~501213). L.W. is supported by the Spanish MICINN (Project No.~FIS2012-36673-C03-03). T.Y. is supported by the NSF PHY-0925174 and DOD/AF/AFOSR No.~FA9550-12-1-0001. L.W. also acknowledges CSRC, Beijing for the hospitality during his visit to it.
\end{acknowledgments}
\appendix

\section{Derivation for the HFD equation with a general bath correlation function}\label{ax_s_gen}

In this appendix, we derive the HFD equation [i.e., Eq.~\eqref{gen_a_qj}] for an open system with a general bath correlation function $\alpha(t,s)$. Define
\begin{align}
	\mathcal{Q}_k^{(\mathbf{j})}(t,z^*)=&\mathcal{P}^{(j_k)}\mathcal{P}^{(j_{k-1})} \ldots \mathcal{P}^{(j_1)}\nonumber\\
	&\times\int_0^t ds_0 \alpha^{(j_0)}(t,s_0) O(t,s_0,z^*),\label{eq_qs_def}
\end{align}
where $\mathbf{j}=\sum_i j_i\mathbf{e_i}\equiv (j_0,j_1,\ldots,j_k)$ with $\mathbf{e_i}$ being the $i$th unit vector, $\mathcal{P}^{(j)}=\int_0^t ds \alpha^{(j)}(t,s) \frac{\delta}{\delta z^*_s}$, and $\alpha^{(j)}(t,s)=\frac{\partial^j}{\partial t^j}\alpha(t,s)$. At the zeroth order, using the quantum state diffusion (QSD) equation
\begin{align}
	\dot{O}(t,s,z^*)]=&\left[-iH_{\rm sys} +Lz_t^*  -L ^\dagger \mathcal{Q}_0^{(0)}(t,z^*),O(t,s,z^*)\right]\nonumber\\
	&-L ^\dagger \frac{\delta \mathcal{Q}_0^{(0)}(t,z^*)}{\delta z_s^*},
\end{align}
we have
\begin{widetext}
\begin{align}
	\frac{\partial}{\partial t}\mathcal{Q}_0^{(j_0)}(t,z^*)&=\frac{\partial}{\partial t}\left[\int_0^t ds_0 \alpha^{(j_0)}(t,s_0) O(t,s_0,z^*)\right] \nonumber\\
	&=\alpha^{(j_0)}(0) O(t,t,z^*)+\int_0^t ds_0 \frac{\partial}{\partial t}\left(\alpha^{(j_0)}(t,s_0)\right) O(t,s_0,z^*)+\int_0^t ds_0 \alpha^{(j_0)}(t,s) \dot{O}(t,s_0,z^*) \nonumber\\
	&=\alpha^{(j_0)}(0) L+\mathcal{Q}_0^{(j_0+1)}(t,z^*)-L ^\dagger \mathcal{Q}_1^{(0,j_0)}(t,z^*)+\left[-iH_{\rm sys} +Lz_t^*  -L ^\dagger \mathcal{Q}_0^{(0)}(t,z^*),\mathcal{Q}_0^{(j_0)}(t,z^*)\right].\label{eq_sp_dq0}
\end{align}
\end{widetext}

For $k\geq 1$, it can be seen that
\begin{align}
	\frac{\partial}{\partial t}\mathcal{P}^{(j)}&=\frac{\partial}{\partial t}\left[\int_0^t ds \alpha^{(j)}(t,s) \frac{\delta}{\delta z_s^*}\right]\nonumber\\
	&=\alpha^{(j)}(0)\frac{\delta}{\delta z_t^*}+\int_0^t ds \alpha^{(j+1)}(t,s) \frac{\delta}{\delta z_s^*}\nonumber\\
	&=\alpha^{(j)}(0)\frac{\delta}{\delta z_t^*}+\mathcal{P}^{(j+1)}.\label{eq_s_ppdot}
\end{align}
Here we define $\mathcal{D}(\mathbf{j},i)\equiv(j_0,\ldots j_{i-1},j_{i+1},\ldots,j_k)$, which excludes $j_i$ in the vector $\mathbf{j}$. For any $n$-dimensional vector $\mathbf{v}=(v_1,\ldots,v_n)$ and integer $x$, the notation $(x,\mathbf{v})$ represents a new $(n+1)$-dimensional vector $\mathbf{v'}=(x,v_1,\ldots,v_n)$. We also introduce a vector $\mathbf{c_i}$ that chooses $i$ elements from $(j_1,\ldots,j_k)$ and its complement $\mathbf{\bar{c}_{i}}$ that contains the remaining elements. For example, for $(1,2,3,4,5)$, $\mathbf{c_2}$ can be $(1,4)$ and then $\mathbf{\bar{c}_{2}}=(2,3,5)$.

We thus have
\begin{align}
	\frac{\partial}{\partial t}\mathcal{Q}_k^{(\mathbf{j})}(t,z^*)&=\frac{\partial}{\partial t}\left[\mathcal{P}^{(j_k)}\mathcal{P}^{(j_{k-1})} \ldots \mathcal{P}^{(j_1)}\mathcal{Q}_0^{(j_0)}(t,z^*)\right]\nonumber\\
	&=\sum_{i=1}^k \mathcal{P}^{(j_k)}\mathcal{P}^{(j_{k-1})} \ldots\frac{\partial \mathcal{P}^{(j_i)}}{\partial t}\ldots \mathcal{P}^{(j_1)}\mathcal{Q}_0^{(j_0)}\nonumber\\
	&\phantom{=}+\mathcal{P}^{(j_k)}\mathcal{P}^{(j_{k-1})} \ldots \mathcal{P}^{(j_1)}\frac{ \partial \mathcal{Q}_0^{(j_0)}(t,z^*)}{\partial t}.\label{eq_s_dq_step0}
\end{align}
Using Eq.~\eqref{eq_s_ppdot} for the first term on the RHS of Eq.~\eqref{eq_s_dq_step0} and Eq.~\eqref{eq_sp_dq0} for the second term, we have
\begin{widetext}
\begin{align}
	\frac{\partial}{\partial t}\mathcal{Q}_k^{(\mathbf{j})}(t,z^*)=&\sum_{i=1}^k \mathcal{P}^{(j_k)}\mathcal{P}^{(j_{k-1})} \ldots\mathcal{P}^{(j_{i+1})}\alpha^{(j_i)}(0)\frac{\delta}{\delta z_t^*}\mathcal{P}^{(j_{i-1})}\ldots \mathcal{P}^{(j_1)}\mathcal{Q}_0^{(j_0)}(t,z^*)\nonumber\\
	&+\sum_{i=1}^k \mathcal{P}^{(j_k)}\mathcal{P}^{(j_{k-1})} \ldots\mathcal{P}^{(j_i+1)}\ldots \mathcal{P}^{(j_1)}\mathcal{Q}_0^{(j_0)}(t,z^*) \nonumber\\
	&+\mathcal{Q}_k^{(j_0+1,j_1,\ldots,j_k)}(t,z^*)+\left[-iH_s +Lz_t^* ,\mathcal{Q}_k^{(j_0,j_1,\ldots,j_k)}(t,z^*)\right]-L ^\dagger \mathcal{Q}_{k+1}^{(0,j_0,j_1,\ldots,j_k)}(t,z^*) \nonumber\\
	&-\mathcal{P}^{(j_k)}\mathcal{P}^{(j_{k-1})} \ldots \mathcal{P}^{(j_1)}\left[L ^\dagger \mathcal{Q}_0^{(0)}(t,z^*),\mathcal{Q}_0^{(j_0)}(t,z^*)\right].
\end{align}
Note that
\begin{align}
	&\mathcal{P}^{(j_k)}\mathcal{P}^{(j_{k-1})} \ldots\mathcal{P}^{(j_i+1)}\ldots \mathcal{P}^{(j_1)}\mathcal{Q}_0^{(j_0)}(t,z^*)=\mathcal{Q}_k^{(\mathbf{j}+\mathbf{e_i})}(t,z^*), \nonumber\\
	&\mathcal{P}^{(j_k)}\mathcal{P}^{(j_{k-1})} \ldots\mathcal{P}^{(j_{i+1})}\alpha^{(j_i)}(0)\frac{\delta}{\delta z_t^*}\mathcal{P}^{(j_{i-1})}\ldots \mathcal{P}^{(j_1)}\mathcal{Q}_0^{(j_0)}(t,z^*)=\alpha^{(j_i)}(0)\left[L,\mathcal{Q}_{k-1}^{(\mathcal{D}
(\mathbf{j},i))}(t,z^*)\right].
\end{align}
Then, we finally have
\begin{align}
	\frac{\partial}{\partial t}\mathcal{Q}_k^{(\mathbf{j})}(t,z^*)&=\sum_{i=1}^k \alpha^{(j_i)}(0)\left[L,\mathcal{Q}_{k-1}^{(\mathcal{D}(\mathbf{j},i))}(t,z^*)\right]+\sum_{i=1}^k \mathcal{Q}_k^{(\mathbf{j}+\mathbf{e_i})}(t,z^*)+\mathcal{Q}_k^{(\mathbf{j}+\mathbf{e_0})}(t,z^*)-L ^\dagger \mathcal{Q}_{k+1}^{(0,\mathbf{j})}(t,z^*)\nonumber\\
	&\phantom{=}+\left[-iH_{\rm sys} +Lz_t^* ,\mathcal{Q}_k^{(\mathbf{j})}(t,z^*)\right]-\mathcal{P}^{(j_k)}\mathcal{P}^{(j_{k-1})} \ldots \mathcal{P}^{(j_1)}\left[L ^\dagger \mathcal{Q}_0^{(0)}(t,z^*),\mathcal{Q}_0^{(j_0)}(t,z^*)\right]\nonumber\\
	&=\sum_{i=1}^k \alpha^{(j_i)}(0)\left[L,\mathcal{Q}_{k-1}^{(\mathcal{D}(\mathbf{j},i))}(t,z^*)\right]+\sum_{i=0}^k \mathcal{Q}_k^{(\mathbf{j}+\mathbf{e_i})}(t,z^*)-L ^\dagger \mathcal{Q}_{k+1}^{(0,\mathbf{j})}(t,z^*)\nonumber\\
	&\phantom{=}+\left[-iH_{\rm sys} +Lz_t^* ,\mathcal{Q}_k^{(\mathbf{j})}(t,z^*)\right]-\sum_{i=0}^k \sum_\mathbf{c_i}\left[L ^\dagger \mathcal{Q}_i^{(0,\mathbf{c_i})}(t,z^*),\mathcal{Q}_{k-i}^{(j_0,\mathbf{\bar{c}_i})}(t,z^*)\right],
\end{align}
\end{widetext}
where $\sum_{\mathbf{c_i}}$ is the sum of $C_i^k\equiv k!/[i!(k-i)!]$ terms that include all possible cases of choosing $i$ elements from a $k$-dimensional vector $(j_1,\ldots,j_k)$. This comes from the term
\begin{equation}
	\mathcal{P}^{(j_k)}\mathcal{P}^{(j_{k-1})} \ldots \mathcal{P}^{(j_1)}\left[L ^\dagger \mathcal{Q}_0^{(0)}(t,z^*),\mathcal{Q}_0^{(j_0)}(t,z^*)\right],
\end{equation}
where we have chosen $i$ operators in $\mathcal{P}^{(j_k)}\mathcal{P}^{(j_{k-1})} \ldots \mathcal{P}^{(j_1)}$ to act on $\mathcal{Q}_0^{(0)}(t,z^*)$, and the remaining $(k-i)$ $\mathcal{P}^{(j_x)}$ operators act on $\mathcal{Q}_0^{(j_0)}(t,z^*)$. For example, for $\mathcal{P}^{(j_5)}\mathcal{P}^{(j_4)} \mathcal{P}^{(j_3)}\mathcal{P}^{(j_2)}\mathcal{P}^{(j_1)}$, one possible case with $i=2$ is that $\mathcal{P}^{(j_3)} \mathcal{P}^{(j_1)}$ acts on $\mathcal{Q}_0^{(0)}(t,z^*)$ to give $\mathcal{Q}_2^{(0,j_1,j_3)}(t,z^*)$, and $\mathcal{P}^{(j_5)}\mathcal{P}^{(j_4)} \mathcal{P}^{(j_2)}$ acts on $\mathcal{Q}_0^{(j_0)}(t,z^*)$ to give $\mathcal{Q}_3^{(j_0,j_2,j_4,j_5)}(t,z^*)$. For the $\mathcal{Q}_k^{(\mathbf{j})}$ operator, termination of the index $k$ is determined by the functional derivative of the $O$ operator, whereas termination of the index $\mathbf{j}$ is determined by the bath correlation function $\alpha(t,s)$. In numerical calculations, this set of equations can be terminated by imposing $\mathcal{Q}_k^{(\mathbf{j})}\equiv 0$ for all $k+\sum_i j_i>\mathcal{N}$, where $\mathcal{N}$ is a suitably chosen integer.

As an example, we use an exactly solvable model to show what the HFD equations look like in the presence of a general bath correlation function. Consider a three-level atom in a multi-mode bosonic bath~\cite{Jing2010_2012}, where $H_{\rm sys}=\omega J_z$, and $L=J_-$. The HFD equation for $\mathcal{Q}_0^{(\mathbf{j})}(t,z^*)$, with $\mathbf{j}\equiv (j_0)$, is given by
\begin{widetext}
\begin{align}
	\frac{\partial}{\partial t}\mathcal{Q}_0^{(j_0)}(t,z^*)&=\alpha^{(j_0)}(0) L+\mathcal{Q}_0^{(j_0+1)}(t,z^*)-L ^\dagger \mathcal{Q}_1^{(0,j_0)}(t,z^*)+\left[-iH_{\rm sys} +Lz_t^*  -L ^\dagger \mathcal{Q}_0^{(0)}(t,z^*),\mathcal{Q}_0^{(j_0)}(t,z^*)\right],\label{hfd_s3a}
\end{align}
where $j_0=0,1,2,\ldots$. The HFD equation for $\mathcal{Q}_1^{(\mathbf{j})}(t,z^*)$, with $\mathbf{j}\equiv(j_0,j_1)$, is given by
\begin{align}
	\frac{\partial}{\partial t}\mathcal{Q}_1^{(j_0,j_1)}(t,z^*)=&\alpha^{(j_1)}(0)\left[L,\mathcal{Q}_{0}^{(j_0)}(t,z^*)\right]+\mathcal{Q}_1^{(j_0+1,j_1)}(t,z^*)+ \mathcal{Q}_1^{(j_0,j_1+1)}(t,z^*)+\left[-iH_{\rm sys} +Lz_t^* ,\mathcal{Q}_1^{(j_0,j_1)}(t,z^*)\right]\nonumber\\
	&-\left[L ^\dagger \mathcal{Q}_0^{(0)}(t,z^*),\mathcal{Q}_{1}^{(j_0,j_1)}(t,z^*)\right]-\left[L ^\dagger \mathcal{Q}_1^{(0,j_0)}(t,z^*),\mathcal{Q}_{0}^{(j_0)}(t,z^*)\right],\label{hfd_s3b}
\end{align}
\end{widetext}
where $j_0,j_1=0,1,2,\ldots$ and $j_0 \leq j_1$. Because the functional expansion of the $O$ operator terminates at the first order for this exactly solvable three-level model, it follows from Eqs.~\eqref{eq_qs_def} and~\eqref{eq_funcexp_def} that
\begin{equation}
	\mathcal{Q}_k^{(\mathbf{j})}(t,z^*)=0,\; \mathbf{j}\equiv (j_0,j_1,\ldots,j_k),\; k\geq 2.
\end{equation}
Solving HFD equations in Eqs.~\eqref{hfd_s3a} and~\eqref{hfd_s3b}, we can obtain $\bar{O}$ for an arbitrary bath correlation function $\alpha(t,s)$ and then determine the quantum-dynamical behavior of the system using the QSD equation. To terminate Eqs.~\eqref{hfd_s3a} and~\eqref{hfd_s3b} in numerical calculations, we can impose $\mathcal{Q}_0^{(j_\mathrm{max}+1)}=a\mathcal{Q}_0^{(j_\mathrm{max})}$ and $\mathcal{Q}_1^{(j_\mathrm{max},j_\mathrm{max}+1)}=a\mathcal{Q}_1^{(j_\mathrm{max},j_\mathrm{max})}$, where $j_\mathrm{max}$ is a suitably chosen integer and $a$ is a parameter determined by the bath correlation function.

In particular, for an Ornstein-Uhlenbeck noise, we have $\mathcal{Q}_0^{(1)}=-\gamma\mathcal{Q}_0^{(0)}=-\gamma\mathcal{Q}_0$ and $\mathcal{Q}_1^{(0,1)}=-\gamma\mathcal{Q}_1^{(0,0)}=-\gamma\mathcal{Q}_1$. Thus, $j_\mathrm{max}=0$ and $a=-\gamma$. Then, Eqs.~\eqref{hfd_s3a} and~\eqref{hfd_s3b} simply reduce to

\begin{align}
	&\frac{\partial}{\partial t}\mathcal{Q}_0(t,z^*)=\alpha(0) L-\gamma\mathcal{Q}_0(t,z^*)-L ^\dagger \mathcal{Q}_1(t,z^*)\nonumber\\
	&+\left[-iH_{\rm sys} +Lz_t^*  -L ^\dagger \mathcal{Q}_0(t,z^*),\mathcal{Q}_0(t,z^*)\right],\label{hfd_s3a_ou}
\end{align}
and
\begin{align}
	&\frac{\partial}{\partial t}\mathcal{Q}_1(t,z^*)=\alpha(0)\left[L,\mathcal{Q}_{0}(t,z^*)\right]-2 \gamma \mathcal{Q}_1(t,z^*)\nonumber\\
	&+\left[-iH_{\rm sys} +Lz_t^* ,\mathcal{Q}_1(t,z^*)\right]-\left[L ^\dagger \mathcal{Q}_0(t,z^*),\mathcal{Q}_{1}(t,z^*)\right]\nonumber\\
	&-\left[L ^\dagger \mathcal{Q}_1(t,z^*),\mathcal{Q}_{0}(t,z^*)\right],\label{hfd_s3b_ou}
\end{align}
which give the results in Ref.~\cite{Jing2010_2012}.

\section{The relationship between SDE and HFD methods}\label{ax_s_rel}

In this appendix, we explicitly show the relationship between the stochastic differential equation (SDE) approach~\cite{Li2014} and our generalized hierarchical functional derivative (HFD) approach.

From the definition of the $Q_m^{(n)}$ operator in Eq.~\eqref{qdef}, we have
\begin{align}
	\int_0^t ds \alpha(t,s)\frac{\delta }{\delta\tilde{z}^*_s}Q_m^{(n)}(t,\tilde{z}^*)=(n-m)Q^{(n)}_{m+1}.\label{eq_ax_qnm}
\end{align}
Thus, from Eqs.~\eqref{eq_qs_def} and~\eqref{fdef}, it follows that
\begin{align}
	\mathcal{Q}_0(t,\tilde{z}^*)\equiv & ~\bar{O}(t,\tilde{z}^*)=\sum_{n=0}^\mathcal{N} Q_0^{(n)}(t,\tilde{z}^*), \nonumber \\
	\mathcal{Q}_1(t,\tilde{z}^*)=&\int ds \alpha(t,s)\frac{\delta }{\delta\tilde{z}^*_s}\mathcal{Q}_{0}(t,\tilde{z}^*) \nonumber\\
	=&\sum_{n=1}^\mathcal{N} n Q_1^{(n)}, \\
	\mathcal{Q}_2(t,\tilde{z}^*)=&\int ds \alpha(t,s)\frac{\delta }{\delta\tilde{z}^*_s}\mathcal{Q}_{1}(t,\tilde{z}^*) \nonumber\\
	=&\sum_{n=2}^\mathcal{N} n (n-1) Q_2^{(n)}, \nonumber\\
	\ldots&\ldots \nonumber
\end{align}
By mathematical induction, it can be obtained that if
\begin{align}
\mathcal{Q}_k(t,\tilde{z}^*)=&\sum_{n=k}^\mathcal{N} n(n-1)\ldots(n-k+1)Q_k^{(n)} \nonumber\\
=&\sum_{n=k}^\mathcal{N} \frac{n!}{(n-k)!}Q_k^{(n)}(t,\tilde{z}^*),
\end{align}
then for $\mathcal{Q}_{k+1}$, we have
\begin{align}
	\mathcal{Q}_{k+1}(t,\tilde{z}^*)=&\sum_{n=k}^\mathcal{N} \frac{n!}{(n-k)!} \int_0^t ds \alpha(t,s)\frac{\delta }{\delta\tilde{z}^*_s}Q_k^{(n)}(t,\tilde{z}^*) \nonumber\\
	=&\sum_{n=k+1}^\mathcal{N} \frac{n!}{(n-k-1)!} Q_{k+1}^{(n)}(t,\tilde{z}^*).
\end{align}
This proves the result given in Eq.~\eqref{fkq}.

When an Ornstein-Uhlenbeck noise is considered, the hierarchical equation of motion for the $Q_k^{(n)}$ operator is derived in Ref.~\cite{Li2014} as
\begin{widetext}
\begin{align}
	\frac{\partial}{\partial t}Q_k^{(n)}(t,\tilde{z}^*)=&\delta_{n,0}\alpha(0)L +\frac{k}{\max\{1,n\}}\alpha(0)\left[L,Q^{(n-1)}_{k-1}(t,\tilde{z}^*)\right]+\frac{(n-k)}{\max\{1,n\}}\tilde{z}_t^* \left[L,Q^{(n-1)}_k(t,\tilde{z}^*)\right] \nonumber\\
	&-(k+1) \gamma Q_{k}^{(n)}(t,\tilde{z}^*)+\left[-iH_{\rm sys},Q_k^{(n)}(t,\tilde{z}^*)\right]-(n+1) L ^\dagger Q^{(n+1)}_{k+1}(t,\tilde{z}^*) \nonumber\\
	&-\sum_{p=0}^{n}\sum_{l} \frac{C^p_l C^{n-p}_{n-k-l}}{C_k^n}\left[L ^\dagger Q^{(p)}_{p-l}(t,\tilde{z}^*),Q^{(n-p)}_{k-p+l}(t,\tilde{z}^*)\right].\label{eq_dqt}
\end{align}
Substituting Eq.~\eqref{eq_dqt} into $\frac{\partial}{\partial t} \mathcal{Q}_k(t,\tilde{z}^*)=\sum_{n=k}^\mathcal{N} \frac{n!}{(n-k)!} \frac{\partial}{\partial t} Q_k^{(n)}(t,\tilde{z}^*)$, it follows that for $k=0$, one has
\begin{align}
	\frac{\partial}{\partial t} \mathcal{Q}_0(t,\tilde{z}^*)=&\sum_{n=0}^\mathcal{N} \frac{\partial}{\partial t} Q_0^{(n)}(t,\tilde{z}^*) \nonumber\\
	=&\alpha(0)L +\sum_{n=0}^\mathcal{N} \bigg\{ \tilde{z}_t^* \left[L,Q^{(n-1)}_0(t,\tilde{z}^*)\right]- \gamma Q_{0}^{(n)}(t,\tilde{z}^*)+\left[-iH_{\rm sys},Q_0^{(n)}(t,\tilde{z}^*)\right]\nonumber\\
	&-\sum_{p=0}^{n}\sum_{l=p}^{p} \frac{C^p_l C^{n-p}_{n-l}}{C_0^n}\left[L ^\dagger Q^{(p)}_{p-l}(t,\tilde{z}^*),Q^{(n-p)}_{l-p}(t,\tilde{z}^*)\right]-(n+1) L ^\dagger Q^{(n+1)}_{1}(t,\tilde{z}^*)\bigg\} \nonumber\\
	=&\alpha(0)L +\tilde{z}_t^* \left[L,\mathcal{Q}_0(t,\tilde{z}^*)\right]- \gamma \mathcal{Q}_{0}(t,\tilde{z}^*)+\left[-iH_{\rm sys},\mathcal{Q}_0(t,\tilde{z}^*)\right] -\left[L ^\dagger \mathcal{Q}_{0}(t,\tilde{z}^*),\mathcal{Q}_{0}(t,\tilde{z}^*)\right]-L ^\dagger \mathcal{Q}_{1}(t,\tilde{z}^*),
\end{align}
and for $k\geq 1$, one has
\begin{align}
	\frac{\partial}{\partial t} \mathcal{Q}_k(t,\tilde{z}^*)=&\sum_{n=k}^\mathcal{N}  \frac{n!}{(n-k)!} \frac{\partial}{\partial t}Q_k^{(n)}(t,\tilde{z}^*) \nonumber\\
	=&\sum_{n=k}^\mathcal{N}\frac{n!}{(n-k)!}\left\{ \frac{k}{n}\alpha(0)\left[L,Q^{(n-1)}_{k-1}(t,\tilde{z}^*)\right]+\frac{(n-k)}{n}\tilde{z}_t^* \left[L,Q^{(n-1)}_k(t,\tilde{z}^*)\right]-(k+1) \gamma Q_{k}^{(n)}(t,\tilde{z}^*) \right.\nonumber\\
	&\left.+\left[-iH_{\rm sys},Q_k^{(n)}(t,\tilde{z}^*)\right]-(n+1) L ^\dagger Q^{(n+1)}_{k+1}(t,\tilde{z}^*)-\sum_{p=0}^{n}\sum_{l} \frac{C^p_l C^{n-p}_{n-k-l}}{C_k^n}\left[L ^\dagger Q^{(p)}_{p-l}(t,\tilde{z}^*),Q^{(n-p)}_{k-p+l}(t,\tilde{z}^*)\right]\right\} \nonumber\\
	=&k\alpha(0)\left[L,\mathcal{Q}_{k-1}(t,\tilde{z}^*)\right]+\tilde{z}_t^* \left[L,\mathcal{Q}_k(t,\tilde{z}^*)\right]-(k+1) \gamma \mathcal{Q}_{k}(t,\tilde{z}^*) - L ^\dagger \mathcal{Q}_{k+1}(t,\tilde{z}^*)\nonumber\\
	&+\left[-iH_{\rm sys},\mathcal{Q}_k(t,\tilde{z}^*)\right]-\sum_{n=k}^\mathcal{N}\frac{n!}{(n-k)!}\sum_{p=0}^{n}\sum_{l} \frac{C^p_l C^{n-p}_{n-k-l}}{C_k^n}\left[L ^\dagger Q^{(p)}_{p-l}(t,\tilde{z}^*),Q^{(n-p)}_{k-p+l}(t,\tilde{z}^*)\right].\label{ax_qk1_im}
\end{align}
The last term on the RHS of Eq.~\eqref{ax_qk1_im} can be rewritten as
\begin{align}
	&\sum_{n=k}^\mathcal{N}\frac{n!}{(n-k)!} \sum_{p,l} \frac{C^p_l C^{n-p}_{n-k-l}}{C_k^n}\left[L ^\dagger Q^{(p)}_{p-l}(t,\tilde{z}^*),Q^{(n-p)}_{k-p+l}(t,\tilde{z}^*)\right] \nonumber\\
	&=\sum_{n=k}\sum_{p,l}  \frac{k!}{(p-l)!(k+l-p)!} \left[L ^\dagger \frac{p!}{l!}Q^{(p)}_{p-l}(t,\tilde{z}^*),\frac{(n-p)!}{(n-k-l)!}Q^{(n-p)}_{k-p+l}(t,\tilde{z}^*)\right]\nonumber\\
	&=\sum_{i} C^k_{i}\left[L ^\dagger \mathcal{Q}_{i}(t,\tilde{z}^*),\mathcal{Q}_{k-i}(t,\tilde{z}^*)\right],\label{ax_cnm}
\end{align}
\end{widetext}
where the substitution $p-l \rightarrow i$ is taken in the last line. Finally, substituting Eq.~\eqref{ax_cnm} into Eq.~\eqref{ax_qk1_im}, we obtain Eq.~\eqref{fk_heq}, i.e., the hierarchical equation of motion for the $\mathcal{Q}_k$ operator. This proves that Eq.~\eqref{eq_dqt} is mathematically equivalent to Eq.~\eqref{fk_heq}.



\end{document}